\documentclass[twocolumn]{aastex631}

\newcommand{\kms}{{\,km\,s$^{-1}$}\xspace} 

\newcommand{\ergs}{{\,erg\,s$^{-1}$}\xspace}
\newcommand{\ergsHz}{{\,erg\,s$^{-1}$\,Hz$^{-1}$}\xspace}
\newcommand{\uJybm}{{\,$\mu$Jy\,beam$^{-1}$}\xspace}
\newcommand{\mJybm}{{\,mJy\,beam$^{-1}$}\xspace}
\newcommand{\pcpc}{{\,\mathrm{pc}^{2}}\xspace}
\newcommand{\masmas}{{\,\mathrm{mas}^{2}}\xspace}

\usepackage{xspace}
\usepackage{color}
\makeatletter
\DeclareRobustCommand{\erase}{\bgroup\markoverwith{\textcolor{red}{\rule[.5ex]{2pt}{1.5pt}}}\ULon}

\makeatother

\submitjournal{Astrophysical Journal}

\begin{document}
\title{
    Origin of Radio Emission in Three Nearby Ultraluminous Infrared Galaxies with Signatures of Luminous Buried Active Galactic Nuclei
}



\correspondingauthor{Takayuki J. Hayashi}
\email{t.hayashi@nao.ac.jp, t.hayashi.bak@gmail.com}

\author[0000-0002-4884-3600]{Takayuki J. Hayashi}
\affiliation{National Astronomical Observatory of Japan \\
2-21-1 Osawa, Mitaka \\
Tokyo 181-8588, Japan}
\affiliation{Azabu Junior and Senior High School\\
2-3-29 Motoazabu, Minato\\
Tokyo 106-0046, Japan}

\author[0000-0002-9043-6048]{Yoshiaki Hagiwara}
\affiliation{Toyo University \\
5-28-20 Hakusan, Bunkyo \\
 Tokyo 112-8606, Japan}

\author[0000-0001-6186-8792]{Masatoshi Imanishi}
\affiliation{National Astronomical Observatory of Japan \\
2-21-1 Osawa, Mitaka \\
Tokyo 181-8588, Japan}
\affiliation{Toyo University \\
5-28-20 Hakusan, Bunkyo \\
 Tokyo 112-8606, Japan}
\affiliation{Department of Astronomy, School of Science, Graduate University
for Advanced Studies (SOKENDAI)\\
2-21-1 Osawa, Mitaka \\
Tokyo 181-8588, Japan}

\begin{abstract}
We report multifrequency Very Long Baseline Array (VLBA) observations at 2.3 and 8.4\,GHz of three nearby ultraluminous infrared galaxies, identified via mid-infrared spectroscopic analyses as hosting deeply embedded active galactic nuclei (AGNs).
Milliarcsecond-scale observations at both frequencies reveal compact continuum emission in IRAS\,F00188$-$0856 and IRAS\,F01298$-$0744, accounting for $\sim10$\% of the flux density measured on arcsecond scales.
The non-detection in IRAS\,F00091$-$0738 and the lower limit on the intrinsic 8.4\,GHz brightness temperature of $10^{6.1}$\,K in IRAS\,F01298$-$0744 yield no conclusive evidence of AGN-driven radio emission, whereas the measurement of $10^{7.8}$\,K in IRAS\,F00188$-$0856  confirms an AGN origin.
Thus, the mid-infrared AGN classification remains robust, with at least one object exhibiting compact radio emission indicative of AGN activity.
We further investigate the high-frequency spectral steepening observed in all three galaxies. 
In each case, this steepening arises from spectral aging in diffuse kpc-scale emission, which is resolved out by the VLBA observations. 
One possible explanation for the steepening of the sample is merger-induced particle acceleration.
IRAS\,F00188$-$0856 exhibits a peaked radio spectrum, characteristic of a young radio source, with the high-frequency steepening attributable to this AGN activity.
Consequently, the spectral steepening at high frequencies arises from particles accelerated by merger dynamics or AGN activity.
\end{abstract}


\section{Introduction} \label{sec:intro}
Ultraluminous infrared galaxies \citep[ULIRGs;][]{1996ARA&A..34..749S} are classified as galaxies with infrared luminosity exceeding $10^{12}L_{\odot}$. 
These systems mainly arise from major mergers, harboring a significant amount of infrared-emitting dust concentrated to their nuclei
\citep{1988ApJ...325...74S,1996MNRAS.279..477C,2000ApJ...529L..77B,2002ApJS..143..315V}, whose powerful energy source arises from the combined contributions of active galactic nuclei (AGNs) and intense starbursts \citep{1998ApJ...498..579G,2009ApJS..182..628V}. 
Unlike optically identified AGNs, enshrouded by toroidal dust \citep[Seyferts;][]{1987ApJS...63..295V,2001ApJS..132...37K,2003MNRAS.346.1055K}, 
compact AGNs postulated to inhabit the nuclei of ULIRGs are enshrouded by dense gas and dust that obscure all lines of sight. 
Consequently, while $\sim 30$\% of ULIRGs display Seyfert-like characteristics in the optical regime \citep{1999ApJ...522..113V}, the detection of AGN signatures is limited by optical observations.
Thus, ULIRGs are presumed to harbor deeply embedded AGNs, concealed from optical observations \citep{2003MNRAS.344L..59M,2006ApJ...637..114I}.

The mid-infrared regime experiences lower dust extinction than optical wavelengths, enabling spectroscopic diagnostics to effectively differentiate obscured AGNs from starburst activity. 
These analyses demonstrate that AGNs play a substantial role in more than half of ULIRGs, even when optical observations do not detect them \citep{2004ApJS..154..178A,2007ApJ...656..148A,2007ApJS..171...72I,2008PASJ...60S.489I,2010ApJ...709..801I,2010ApJ...721.1233I,2009ApJ...694..751I,2008MNRAS.385L.130N,2009MNRAS.399.1373N,2010MNRAS.405.2505N,2009ApJS..182..628V}.
However, highly concentrated starbursts in the central regions may exhibit infrared spectra that resemble those of the buried AGN \citep{2004A&A...414..873S}. 
Although this counterargument is considered physically extreme \citep{2000AJ....119..509S,2005ApJ...630..167T,2007ApJS..171...72I}, the energetic contribution of AGNs remains inconclusive. 
Furthermore, submillimeter observations have reported elevated HCN-to-HCO$^{+}$ flux ratios, indicating the presence of deeply buried AGNs in ULIRGs, even in the absence of mid-infrared AGN signatures \citep{2016AJ....152..218I,2018ApJ...856..143I,2019ApJS..241...19I,2023ApJ...954..148I}. 
These findings underscore the limitations of relying solely on mid-infrared observations to probe ULIRG's nuclei. 
Complementary observations at other wavelengths, such as radio, where dust opacity is likewise diminished, are essential.

Radio emission of ULIRGs is, in general, a mixture of those originating from thermal and nonthermal plasma, 
where the former originates from star formation while the latter is also related to AGNs.
Despite the possible contribution of AGN activity to the radio emission of ULIRGs, previous studies occasionally underestimated its presence in the derivation of various physical quantities \citep[e.g.,][]{2003A&A...409...99P,2006ApJ...645..186T,2010MNRAS.405..887C}.
Their premise is that star formation produces the correlation between radio and far-infrared luminosities of galaxies \citep{1991ApJ...378...65C,2001ApJ...554..803Y}.
However, radio-quiet AGNs show a similar relation \citep{2010ApJ...724..779M}, and hence radio emission of ULIRGs is not necessarily dominated by starbursts \citep{2021A&ARv..29....2P}.
Thus, the origin of the radio emission characteristics in ULIRGs remains enigmatic, necessitating further investigation at radio wavelengths.

To validate the previous diagnosis of AGNs by discussing the origin of the radio spectra of ULIRGs, 
\citet[hereafter H21]{2021MNRAS.504.2675H} have conducted multifrequency observations of ULIRGs whose buried AGNs are not found at optical wavelengths but identified at other wavelengths \citep{2006ApJ...637..114I,2007ApJS..171...72I,2019ApJS..241...19I,2014AJ....148....9I}.
As a result, their radio properties are statistically similar to that of the entire ULIRG sample in \citet{2008A&A...477...95C} \citep[see also][]{2013ApJ...777...58M}.
Although typical ULIRGs exhibit steep nonthermal spectra with flattening at high frequencies due to contribution of free-free emission (FFE) by thermal plasma \citep{1991ApJ...378...65C,2016MNRAS.461..825G,2018MNRAS.474..779G}, certain sources indicate steepening at high frequencies attributed to spectral aging of nonthermal plasma \citep{2008A&A...477...95C,2010MNRAS.405..887C,2011ApJ...739L..25L,2013ApJ...777...58M,2021MNRAS.504.2675H,2025PASA...42....2G}. 
\citet{2011ApJ...739L..25L} have discussed the potential for identifying a spectral break as a means of distinguishing AGNs from starbursts (see also H21). 
On the other hand, \cite{2013ApJ...777...58M} have reported the impact of merger activity, where the radio spectrum steepens at high frequencies as the merger progresses. 
To determine the origin of the steepening, it is crucial to examine the contribution of high-brightness radio sources inherent to an AGN through observations with high angular resolution.

This paper presents the results of multifrequency radio observations conducted using the Very Long Baseline Array (VLBA). 
The high resolution achieved by very long baseline interferometry (VLBI) offers valuable insights into the presence of an AGN behind the dust responsible for infrared and submillimater radiation from ULIRGs.
Although VLBI imaging observations of ULIRGs have been widely conducted, they have been confined to a limited number of objects (e.g., \citealt{1998AJ....115..928C,2000ApJ...532L..95C}, \citealt{1998ApJ...493L..17S}, \citealt{2003ApJ...587..160M,2006ApJ...653.1172M},
\citealt{2005ApJ...618..705P},
\citealt{2012MNRAS.422.1453N},
\citealt{2012MNRAS.422..510R},
\citealt{2019A&A...623A.173V},
\citealt{2024A&A...687A.193W},
\citealt{2024ApJ...970....5H}, and references therein). 
These studies did not target elusive AGNs identified indirectly through infrared observations, nor did they attempt to elucidate the origin of high-frequency steepening. 
The new VLBA observations of three ULIRGs exhibiting mid-infrared AGN signatures and high-frequency spectral steepening offer critical insight into the physical mechanisms governing radio emission in ULIRGs.
This study reports the results of VLBA observations of ULIRGs selected based on these criteria.

In this research, we used the standard cosmological model with cold dark matter and a cosmological constant, adopting $H_0 = 70$\,\kms\,Mpc$^{-1}$, $\Omega_{\rm M} = 0.3$, and $\Omega_\Lambda = 0.7$, supported by observational studies from the past decades \citep[e.g.,][]{2020A&A...641A...6P}.
The paper defines a spectral index, $\alpha$, as $S_\nu \propto \nu^\alpha$, where $S_\nu$ represents the flux density at frequency, $\nu$. 

\section{Sample} \label{sec:sample}
We selected three ULIRGs --- IRAS\,F00091$-$0738, IRAS\,F00188$-$0856, and IRAS\,F01298$-$0744 --- as the targets for this investigation (hereafter referred to as F00091$-$0738, F00188$-$0856, and F01298$-$0744, respectively).
A summary of our sample is provided in Table\,\ref{tbl:sample}.
These ULIRGs exhibit a change in radio spectral index around $\sim$\,10\,GHz, accompanied by a flux deficit at higher frequencies (H21).
The original sample of H21 is derived from ULIRGs in the IRAS 1\,Jy sample \citep{1998ApJS..119...41K}.
H21 conducted multifrequency radio observations using the Karl G. Jansky Very Large Array (VLA) for ULIRGs that lack optical signatures of AGNs \citep{1999ApJ...522..113V}, yet exhibit mid-infrared signatures of obscured AGNs \citep{2006ApJ...637..114I, 2007ApJS..171...72I}.
Thus, their optical classification is characterized as LINER (Low-Ionization Nuclear Emission-Line Region) or \ion{H}{2}.
Although the time sampling is still inadequate for definitively concluding flux-density stability, H21 reported that these sources exhibit no statistically significant variability across the three-epoch dataset at 1.4\,GHz acquired with the JVLA.

Given the correlation between radio emission in ULIRGs and merger activity \citep{2013ApJ...777...58M}, the merger phase of each target represents a critical parameter.
The merger stages of the sample delineated by \cite{2002ApJS..143..315V}, based on optical and near-infrared observations \citep{2002ApJS..143..277K} are as follows: 
F00091$-$0738 resides in the pre-merger phase, comprising a close pair with a nuclear separation of 2.1\,kpc; 
F00188$-$0856 is classified as an old merger, characterized by a single nucleus and exhibiting subtle morphological distortions; 
F01298$-$0744 represents an ongoing merger, characterized by a distorted morphology and two prominent tidal tails, indicative of recent dynamical interaction.

\begin{deluxetable*}{cccccccccccc}
\renewcommand{\tabcolsep}{1mm}
\tabletypesize{\scriptsize}
\tablewidth{0pt} 
\tablecaption{The ULIRG Sample for the VLBA Observations. \label{tbl:sample}}
\tablehead{																					
\colhead{}	&	\multicolumn{2}{c}{Optical Position}			&	\colhead{}	&	\colhead{}	&	\colhead{}	&	\colhead{}	&	\colhead{}	&	\colhead{}	&	\multicolumn{2}{c}{Submillimeter Position}			\\\cline{2-3}\cline{10-11}
\colhead{Object}	&	\colhead{R.A.}	&	\colhead{decl.}	&	\colhead{$z$}	&	\colhead{Linear Scale}	&	\colhead{$\log L_\mathrm{IR}$}	&	\colhead{Optical Class}	&	\colhead{Merger Stage}	&	\colhead{$S_{9.0}$}	&	\colhead{R.A.}	&	\colhead{decl.}	\\
\colhead{}	&	\colhead{}	&	\colhead{}	&	\colhead{}	&	\colhead{(pc\,mas$^{-1}$)}	&	\colhead{($L_\odot$)}	&	\colhead{}	&	\colhead{}	&	\colhead{(mJy\,beam$^-1$)}	&	\colhead{}	&	\colhead{}	\\
}																					
\startdata																					
F00091$-$0738	&	$00^\mathrm{h}11^\mathrm{m}43\fs25$	&	$-07\arcdeg22\arcmin07\farcs5$	&	0.118	&	2.132	&	12.19	&	\ion{H}{2}	&	pre-merger	&	$3.0\pm0.2$	&	$00^\mathrm{h}11^\mathrm{m}43\fs273$	&	$-07\arcdeg22\arcmin07\farcs35$	\\
F00188$-$0856	&	$00^\mathrm{h}21^\mathrm{m}26\fs48$	&	$-08\arcdeg39\arcmin27\farcs1$	&	0.128	&	2.287	&	12.33	&	LINER	&	old merger	&	$4.3\pm0.2$	&	$00^\mathrm{h}21^\mathrm{m}26\fs514$	&	$-08\arcdeg39\arcmin26\farcs01$	\\
F01298$-$0744	&	$01^\mathrm{h}32^\mathrm{m}21\fs41$	&	$-07\arcdeg29\arcmin08\farcs9$	&	0.136	&	2.408	&	12.27	&	\ion{H}{2}	&	ongoing merger	&	$3.2\pm0.2$	&	$01^\mathrm{h}32^\mathrm{m}21\fs413$	&	$-07\arcdeg29\arcmin08\farcs34$	\\
\enddata			
\tablecomments{The optical position, redshift ($z$), and infrared bolometric luminosity ($L_\mathrm{IR}$) are taken from \cite{1998ApJS..119...41K}. 
The optical classification and merger stage are identified by \cite{1999ApJ...522..113V} and \cite{2002ApJS..143..315V}, respectively.
The integrated flux density at 9.0\,GHz ($S_{9.0}$) measured with the VLA, is reported by \cite{2021MNRAS.504.2675H}. 
The submillimeter position is obtained from \cite{2019ApJS..241...19I}.}
\end{deluxetable*}

\section{Observations} \label{sec:obs}
The VLBA observations described in this paper were conducted in May 2022 under project code BH237. 
This paper presents results from three 8\,hour sessions: sessions A, B, and D. 
The results of session C appear in \cite{2024ApJ...970....5H}.
A summary of the observations is provided in Table\,\ref{tbl:obssum}.

During observations, four 128\,MHz frequency subbands were employed for both the right and left circular polarizations.
Data were sampled at the Nyquist rate with two bits per sample, yielding a total data rate of 4\,Gbit\,s$^{-1}$. 
Only parallel polarization products were processed using the DiFX correlator \citep{2011PASP..123..275D}.
Simultaneous observations were made in the $S$ and $X$ bands. 
The $S$-band observations were centered at 2.268\,GHz (hereafter 2.3\,GHz) with a bandwidth of 128\,MHz, while the $X$-band observations comprised subbands centered at 8.304, 8.432, and 8.560\,GHz, with a central frequency of 8.432\,GHz (hereafter 8.4\,GHz) and a total bandwidth of 384\,MHz. 
Each 128\,MHz subband was divided into 256 spectral channels to mitigate radio frequency interference (RFI).
We note that 8.4\,GHz is currently the highest frequency of VLBA data available for these sources.

Observations included scans of S5\,0016$+$73 or BL\,Lac, serving as a fringe finder and bandpass calibrator.
Target data were acquired in phase-reference mode, with 3--4 minute scans on the targets interleaved with calibrator scans. 
The calibrators and their typical scan durations were as follows:
20\,seconds for J0006$-$0623 located 1\fdg7 from the target F00091$-$0738; 
90\,seconds for J0024$-$0811 located 0\fdg8 from the target F00188$-$0856; and 
120\,seconds for J0132$-$0804 located 0\fdg6 from the target F01298$-$0744.
The accurate radio coordinates of the calibrators are provided by the third realization of the International Celestial Reference Frame \citep[ICRF3;][]{2020A&A...644A.159C}.
During target scans, both antennas and correlator were aligned with the optical positions of the targets listed in Table\,\ref{tbl:sample}.

\section{Data Reduction}\label{sec:reduction}
We reduced the data utilizing the Astronomical Image Processing System \citep[AIPS;][]{2003ASSL..285..109G}, developed by the National Radio Astronomy Observatory (NRAO). 
All analyzes followed a standardized procedure and were performed independently for the 2.3 and 8.4\,GHz bands. 
To minimize the influence of atmospheric propagation effects, data acquired at elevation angles below $15\arcdeg$ were flagged. 
Digital sampler bias corrections were applied through the task ACCOR, followed by \textit{a priori} amplitude calibration via the task APCAL, incorporating system noise temperature and station-specific gain values. 
During this phase, no atmospheric opacity correction was implemented. 
Earth orientation parameters and ionospheric dispersive delays were corrected for the 2.3 and 8.4\,GHz data using the tasks VLBAEOPS and VLBATECR, respectively. 
Fringe fitting for calibrators was performed on scan-integrated data using the task FRING. 
For the 8.4\,GHz data, subband delay solutions were derived with the task MBDLY. 
Bandpass calibration for amplitude and phase was applied using the task BPASS, employing a fringe finder as the reference. 
For the 2.3\,GHz data, RFI was identified and iteratively flagged in both the time and frequency domains using the tasks WIPER and SPFLG at each processing stage.

Imaging was performed utilizing the CLEAN algorithm within the Difmap software \citep{1997ASPC..125...77S}. 
The visibilities of the phase calibrators were initially averaged over 20\,second intervals. 
Subsequently, iterative CLEAN and phase-only self-calibration were performed, progressively reducing the solution time intervals to 120, 60, and 30\,minutes. 
Then, amplitude self-calibration was conducted, reducing the solution intervals to 120, 60, and 30 minutes. 
At each stage, the coherence of the gain solutions was meticulously verified. 
Finally, antenna-based gain corrections for the RR and LL visibilities were independently derived using the AIPS task CALIB \citep[see][]{EVNmemo78}. 
These gain solutions, in conjunction with the delay and rate solutions obtained from fringe fitting, were applied to the target data via the AIPS task CLCAL.

The structure and brightness of the phase calibrators in the resulting uniformly weighted images are described as follows:
J0006$-$0623 exhibits a core-jet morphology elongated toward the west at both 2.3 and 8.4\,GHz.
Its peak and integrated flux densities 
at 2.3\,GHz are 1124\mJybm and 1737\,mJy, respectively, 
while at 8.4\,GHz, they are 2407\mJybm and 3544\,mJy, respectively.
J0024$-$0811 appears point-like at 2.3\,GHz and 
shows a core-jet morphology elongated toward the south at 8.4\,GHz.
Its peak and integrated flux densities 
at 2.3\,GHz are 510\mJybm and 563\,mJy, respectively, 
whereas at 8.4\,GHz, they are 418\mJybm and 537.1\,mJy, respectively.
J0132$-$0804 exhibits a two-sided morphology elongated along the east-west direction at both 2.3 and 8.4\,GHz.
Its peak and integrated flux densities 
at 2.3\,GHz are 110\mJybm and 41\,mJy, respectively, 
while at 8.4\,GHz, they are 232\mJybm and 100\,mJy, respectively.

This study investigates whether dust in the central regions of ULIRGs, which emit infrared and submillimeter radiation, is predominantly heated by AGN activity. 
To address this question, high-resolution imaging was performed around the submillimeter continuum source, identified in ALMA observations at $\sim240$\,GHz \citep[see Table\,\ref{tbl:sample};][]{2019ApJS..241...19I}.
After realigning the phase center of the target visibility data to the position of the submillimeter continuum source using the AIPS task UVFIX, the subsequent imaging was performed with the Difmap software.

In the Difmap software, we initially integrated the visibilities over 30\,second intervals. 
Afterward, wide-field maps with field of view of $\sim$\,1\farcs6 at 2.3\,GHz and $\sim$\,0\farcs6 at 8.4\,GHz were generated, utilizing various UV tapering configurations. 
This field of view exceeds the full-width at half maximum (FWHM) of the submillimeter continuum sources reported in \cite{2019ApJS..241...19I,2023ApJ...954..148I}.
In these preliminary dirty maps, we tried to identify radio sources with a signal-to-noise ratio of 5 or greater. 
The imaging was performed by iterative CLEAN and phase-only self-calibration, with the solution time interval initially set to 240\,minutes, and, in some instances, progressively reduced to 120\,minutes. 
For diffuse sources detectable solely by a UV tapered map, imaging was executed without the application of self-calibration.

Throughout the observations conducted in the 8.4\,GHz band, the amplitude fluctuations, as estimated through self-calibration for the calibrators, consistently aligned with the absolute flux density uncertainties of 5\% for the VLBA at 15\,GHz \citep{2002ApJ...568...99H}, a value also applied to the 8.4\,GHz data \citep[e.g.,][]{2020ApJ...891...59R}.
For the 2.3\,GHz band, following the observations in 2022, the VLBA Observational Status Summary for Semester 2024A\footnote{https://science.nrao.edu/facilities/vlba/docs/manuals/oss2024A} has reported  that RFI contamination degrades the system noise temperature, thereby undermining the accuracy of flux calibration and that, even with precise calibration, the flux uncertainty remains no better than 50\% \citep{VLBAmemo41}.
Although we present the 2.3\,GHz flux measurements adopting this uncertainty value tentatively, 
it serves only as an indicative value, given that the observations lacked the strategy required for accurate flux scaling and robust uncertainty assessment.
As the status summary emphasizes, the derived flux density is not reliable enough for scientific analysis or spectral index estimation.
Therefore, the 2.3\,GHz data in this study only serve to enhance the robustness of the 8.4\,GHz detection.

\movetabledown=50mm
\begin{rotatetable*}
\begin{deluxetable*}{ccccccccccccccc}
\tabletypesize{\scriptsize}
\tablewidth{0pt} 
\tablecaption{The VLBA Observations Summary. \label{tbl:obssum}}
\tablehead{					
\colhead{Target}	&	\colhead{Session}	&	\colhead{Date}	&	\colhead{Fringe}  	&	\multicolumn{3}{c}{Phase Calibrator}					&	\colhead{Separation}	&	\colhead{Integration} 	&	\colhead{Band}	&	\colhead{Central}	&	\colhead{Band}	&	\colhead{Missing}	\\\cline{5-7}
\colhead{}	&	\colhead{} 	&	\colhead{}	&	\colhead{Finder}  	&	\colhead{Name}  	&	\colhead{R.A.}	&	\colhead{decl.}	&	\colhead{}	&	\colhead{Time} 	&	\colhead{}	&	\colhead{Frequency}	&	\colhead{Width}	&	\colhead{Station}	\\
\colhead{}	&	\colhead{}	&	\colhead{}	&	\colhead{}	&	\colhead{}	&		&		&	\colhead{(deg)}	&	\colhead{(minutes)}	&	\colhead{}	&	\colhead{(GHz)}	&	\colhead{(MHz)}	&	\colhead{}	\\
}																												
\startdata												
F00091$-$0738	&	A	&	2022 May 7 	&	BL\,Lac	&	J0006$-$0623	&	$00^{\mathrm h}06^{\mathrm m}13\fs892891$	&	$-06\arcdeg23\arcmin35\farcs33532$	&	1.68	&	265	&	$S$	&	2.268	&	128	&	PT	\\
	&		&		&		&		&		&		&		&		&	$X$	&	8.432	&	384	&	FD, PT	\\
F00188$-$0856	&	B	&	2022 May 31	&	BL\,Lac	&	J0024$-$0811	&	$00^{\mathrm h}24^{\mathrm m}00\fs672759$	&	$-08\arcdeg11\arcmin10\farcs04863$	&	0.79	&	216	&	$S$	&	2.268	&	128	&	HN, LA	\\
	&		&		&		&		&		&		&		&		&	$X$	&	8.432	&	384	&	HN, LA	\\
F01298$-$0744	&	D	&	2022 May 30	&	S5\,0016$+$73	&	J0132$-$0804	&	$01^{\mathrm h}32^{\mathrm m}41\fs126049$	&	$-08\arcdeg04\arcmin04\farcs83684$	&	0.59	&	217	&	$S$	&	2.268	&	128	&	BR, LA, HN	\\
	&		&		&		&		&		&		&		&		&	$X$	&	8.432	&	384	&	LA, HN	\\
\enddata																
\end{deluxetable*}
\end{rotatetable*}

\section{Results} \label{sec:res}
The properties of the images obtained for each object, along with the flux measurement results for compact radio sources, are summarized in Tables~\ref{tbl:image} and \ref{tbl:flux}, respectively. 
A detailed explanation of each object is provided below.

\begin{deluxetable}{cccccccccccc}
\tabletypesize{\scriptsize}
\tablewidth{0pt} 
\tablecaption{Synthesized beam and rms noise level of the VLBA images. \label{tbl:image}}
\tablehead{									
\colhead{}	&	\colhead{}	&	\multicolumn{2}{c}{Beam Size}			&	\colhead{}	\\\cline{3-4}
\colhead{Target}	&	\colhead{Band}	&	\colhead{FWHM}	&	\colhead{PA}	&	\colhead{rms}	\\
\colhead{}	&	\colhead{}	&	\colhead{}	&	\colhead{}	&	\colhead{}	\\
\colhead{}	&	\colhead{(GHz)}	&	\colhead{(mas$^2$)}	&	\colhead{(deg)}	&	\colhead{($\mu$Jy\,beam$^{-1}$)}	\\
} 									
\startdata 									
F00091$-$0738	&	2.3	&	$8.3\times3.0$	&	\phn0	&	\phn59	\\
	&	8.4	&	$2.0\times0.9$	&	\phn3	&	\phn24	\\
F00188$-$0856	&	2.3	&	$7.7\times2.7$	&	\phn1	&	\phn88	\\
	&	8.4	&	$2.6\times1.1$	&	$13$	&	\phn31	\\
F01298$-$0744	&	2.3	&	$8.4\times2.7$	&	\phn2	&	178	\\
	&		&	$20\times11$\tablenotemark{\footnotesize a}	&	$13$\tablenotemark{\footnotesize a}	&	404\tablenotemark{\footnotesize a}	\\
	&	8.4	&	$2.5\times1.1$	&	$11$	&	\phn27	\\
	&		&	$3.8\times2.6$\tablenotemark{\footnotesize b}	&	$12$\tablenotemark{\footnotesize b}	&	\phn27\tablenotemark{\footnotesize b}	\\
\enddata    									
\tablecomments{
The measurements were derived from naturally weighted images. 
}
\tablenotetext{a}{The values on the tapered image, utilizing a Gaussian with a FWHM of 10\,M$\lambda$.}
\tablenotetext{b}{The values on the tapered image, utilizing a Gaussian with a FWHM of 50\,M$\lambda$.}
\end{deluxetable}

\begin{deluxetable*}{cccccccccccc}
\tabletypesize{\scriptsize}
\tablewidth{0pt} 
\tablecaption{Flux Measurement of the Targets \label{tbl:flux}}
\tablehead{																			
\colhead{}	&	\colhead{}	&	\colhead{} 	&	\colhead{} 	&	\multicolumn{2}{c}{Source Size}			&	\colhead{} 	&	\multicolumn{2}{c}{Deconvolved Source Size}			&	\colhead{} 	\\\cline{5-6}\cline{8-9}
\colhead{Target}	&	\colhead{Band}	&	\colhead{Peak} 	&	\colhead{Integrated} 	&	\colhead{FWHM} 	&	\colhead{PA} 	&  	\colhead{$T_\mathrm{b}$} 	&	\colhead{FWHM} 	&	\colhead{PA} 	&  	\colhead{$T_\mathrm{b}'$} 	\\
\colhead{}	&	\colhead{}	&	\colhead{Flux Density} 	&	\colhead{Flux Density} 	&	\colhead{}	&	\colhead{}	&  	\colhead{} 	&	\colhead{}	&	\colhead{}	&  	\colhead{} 	\\
\colhead{}	&	\colhead{(GHz)}	&	\colhead{($\mu$Jy\,beam$^{-1}$)} 	&	\colhead{($\mu$Jy)}	&	\colhead{(mas$^2$)} 	&	\colhead{(deg)} 	&   	\colhead{(K)}  	&	\colhead{(mas$^2$)} 	&	\colhead{(deg)} 	&   	\colhead{(K)}  	\\
} 			
\startdata 	
F00091$-$0738	&	2.3	&	$<177$	&	$<177$	&	…	&	…	&	$<10^{6.4}$	&	…	&	…	&	…	\\
	&	8.4	&	$<\phn72$	&	$<\phn72$	&	…	&	 …	&	$<10^{6.1}$	&	 …	&	 …	&	 …	\\
F00188$-$0856	&	2.3	&	$1002\pm\phn509$	&	\phn$831\pm\phn417$	&	\phn$8.2\pm 0.7 \times \phn3.3 \pm  0.3$	&	$176 \pm \phn4$	&	$10^{7.2}$	&	$3.2^{+1.5}_{-3.2} \times 1.2^{+1.8}_{-1.2}$	&	$ 147^{+38}_{-45}$	&	$10^{8.0}$	\\
	&		&		&		&	(\phn$19\pm 2 \times \phn7.5 \pm  0.7$)	&		&		&	($7.3^{+3.4}_{-7.3} \times 2.7^{+4.1}_{-2.7}$)	&		&		\\
	&	8.4	&	\phn$284\pm\phn\phn34$	&	\phn$303\pm\phn\phn37$	&	\phn$3.0\pm 0.3 \times \phn1.2 \pm  0.1$	&	\phn$21 \pm \phn4$	&	$10^{6.4}$	&	$1.6^{+0.5}_{-0.7} \times 0.08^{+0.67}_{-0.08}$	&	$36^{+15}_{-27}$	&	$10^{7.8}$	\\
	&		&		&		&	(\phn$6.8\pm 0.7 \times \phn2.7 \pm  0.2$)	&		&		&	($3.7^{+1.1}_{-1.6} \times 0.2^{+1.5}_{-0.2}$)	&		&		\\
F01298$-$0744	&	2.3	&	$2081\pm1116$	&	$2740\pm1567$	&	$21\pm 3 \times 20 \pm  3$	&	\phn$13 \pm 90$	&	$10^{6.3}$	&	$16^{+5}_{-16} \times 7^{+13}_{-7}$	&	$102\pm14$	&	$10^{7.0}$	\\
	&		&		&		&	($51\pm 7 \times 48 \pm  7$)	&		&		&	($39^{+12}_{-39} \times 17^{+31}_{-17}$)	&		&		\\
	&	8.4	&	\phn$163\pm\phn\phn28$	&	\phn$201\pm\phn\phn33$	&	\phn$3.8\pm 0.6 \times \phn2.9 \pm  0.5$	&	\phn$20 \pm 24$	&	$10^{5.7}$	&	$1.6^{+1.0}_{-1.6} \times <1.6$	&	$171^{+43}_{-50}$	&	$>10^{6.1}$	\\
	&		&		&		&	(\phn$9.2\pm 1.4 \times \phn6.9 \pm  1.2$)	&		&		&	($3.8^{+2.4}_{-3.8} \times <3.8$)	&		&		\\
\enddata    	 				
\tablecomments{All values are measured in the naturally weighted images. 
The flux density error is determined by taking the rms of the thermal noise and the systematic uncertainty of the amplitude calibration, set at 50\% for 2.3\,GHz and 5\% for 8.4\,GHz. 
For F01298$-$0744, the images for the flux measurements were tapered using a Gaussian with an FWHM of 10\,M$\lambda$ at 2.3\,GHz and 50\,M$\lambda$ at 8.4\,GHz.
For each source detected at each frequency for each object, the deconvolved source size derived from the AIPS task IMFIT is presented, along with the corresponding intrinsic brightness temperature ($T_\mathrm{b}'$) calculated from these parameters.
For both the observed and deconvolved source sizes, the bracketed values denote the projected linear size in units of pc$^2$.
}
\end{deluxetable*}

\subsection{F00091\texorpdfstring{$-$}{-}0738}
The full-resolution image at 2.3\,GHz, without UV tapering, exhibits a thermal noise level of 59\uJybm and a beam size of $8.3 \times 3.0\masmas$ ($18 \times 6.4\pcpc$). 
At 8.4\,GHz, the thermal noise level is 24\uJybm with a beam size of $2.0 \times 0.87\masmas$ ($4.2 \times 1.9\pcpc$). 
No radio source was detected at the location of the submillimeter continuum source in either image, even after the UV tapering.
We adopted a $3\sigma$ threshold as the upper limit for the flux density of any potential radio source in full-resolution images, resulting in surface brightness limits of 177\uJybm at 2.3\,GHz and 72\uJybm at 8.4\,GHz (Table\,\ref{tbl:flux}). 
Using these values, we derived the upper limit for the brightness temperature, $T_\mathrm{b}$ (in K), with the following equation:
\begin{eqnarray}
T_{\mathrm b} =  
1.8\times 10^{9} (1+z)
S_\nu
\nu^{-2}
\phi^{-2}
\label{eq:Tb},
\end{eqnarray}
where $S_\nu$ and $\phi$ are the flux density (in mJy) at frequency, $\nu$ (in GHz), and source size (in mas), respectively \citep[see][]{2005ApJ...621..123U}.
Consequently, the upper limits of the brightness temperature are $10^{6.4}$\,K at 2.3\,GHz and $10^{6.1}$\,K at 8.4\,GHz for compact radio sources behind the dust emitting submillimeter and infrared radiation.

\subsection{F00188\texorpdfstring{$-$}{-}0856}
Full-resolution images without UV tapering achieved a beam size of $7.7 \times 2.7\masmas$ ($18 \times 6.2\pcpc$) at 2.3\,GHz, 
and $2.6 \times 1.1\masmas$ ($5.9 \times 2.5$ pc$^2$) at 8.4\,GHz (Table\,\ref{tbl:image}).
In the images at both frequencies, we detected a point source exceeding 5$\sigma$ threshold of the thermal noise.
The flux peak is located at
$\mathrm{R.A.} = 00^{\mathrm h}21^{\mathrm m}26\fs51252 \pm 0\fs00006$ and 
$\mathrm{decl.} = -08\arcdeg39\arcmin25\farcs995 \pm 0\farcs002$ at 2.3\,GHz, and 
$\mathrm{R.A.} = 00^{\mathrm h}21^{\mathrm m}26\fs51258 \pm 0\fs00003$ and 
$\mathrm{decl.} = -08\arcdeg39\arcmin25\farcs994 \pm 0\farcs001$ at 8.4\,GHz.
Position error estimates are detailed in Appendix \ref{sec:position}.
The positional offset of the radio sources between the two frequencies amounts to 0.7$\sigma$.
This radio source is unlikely to be noise, as signals above the threshold at different frequencies, which we analyzed separately, were detected at positions that match within the error margin.

\begin{figure*}[ht]
    \centering
    \includegraphics[width=0.8\linewidth]{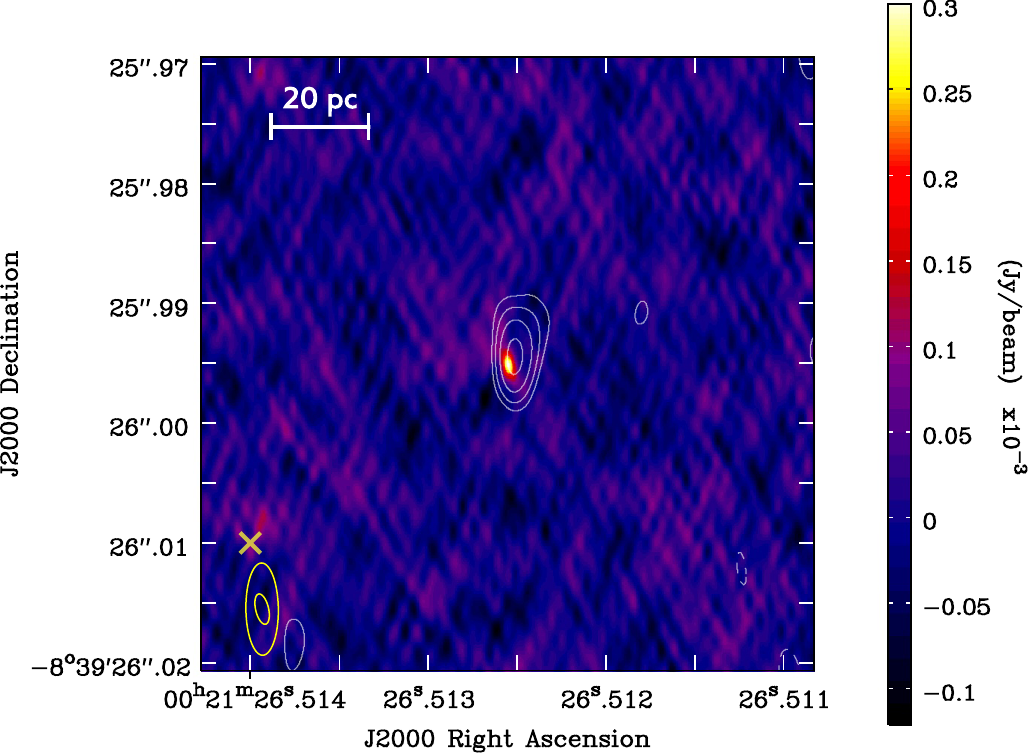}
    \caption{
        A naturally weighted VLBA image of F00188$-$0856 at 8.4\,GHz with 2.3 GHz contours overlaid.
        Contour levels begin at 264\uJybm, corresponding to the 3$\sigma$ noise level of the image, and increase by factors of $\sqrt{2}$. 
        The resolutions at 2.3 and 8.4\,GHz are $7.7\times2.7\masmas$ at a PA of 0\fdg7 and $2.6\times1.1\masmas$ at a PA of 13\fdg2, respectively, indicated by the yellow ellipses in the bottom left of the image.
        A yellow cross indicates the location of the submillimeter continuum source detected by ALMA observations \citep{2019ApJS..241...19I}, which corresponds to the phase center in imaging process.
        The peak flux densities are 831\uJybm and 303\uJybm at 2.3 and 8.4\,GHz, respectively.
        \label{fig:F00188img}
        }
\end{figure*}

Figure\,\ref{fig:F00188img} shows an image of F00188$-$0856 after phase-only self-calibration. 
The thermal noise and peak flux density measured in the image are 88 and 831\uJybm at 2.3\,GHz and 31 and 303\uJybm at 8.4\,GHz, respectively.
The Gaussian fit for the image performed using the AIPS task IMFIT yields a FWHM of the source as $8.2\pm 0.7 \times 3.3 \pm  0.3\masmas$ in the 2.3\,GHz image, whereas $3.0\pm 0.3 \times 1.2 \pm  0.1\masmas$ in the 8.4\,GHz image  (Table\,\ref{tbl:image}).
We estimated the flux density of the source by summing the CLEAN components within these regions, producing a value of 1002 and 284\,$\mu$Jy at 2.3 and 8.4\,GHz, respectively\footnote{
    The Difmap software computes the peak flux density by converting the flux density per pixel in the image into flux density per beam size. 
    Consequently, the flux density value of an unresolved source, smaller than the beam size, can be less than its peak flux density value.
} (Table\,\ref{tbl:flux}).

The brightness temperature estimates, based on the observed source sizes at 2.3 and 8.4\,GHz, yield values of $T_\mathrm{b}\sim10^{7.2}$\,K and $\sim10^{6.4}$\,K, respectively, using Equation\,(\ref{eq:Tb}). 
However, these values inherently depend on the beam size of the observations. 
To refine the estimates, we used the deconvolved sizes derived from the IMFIT, measured as 
$3.2^{+1.5}_{-3.2} \times 1.2^{+1.8}_{-1.2}\masmas$ and 
$1.6^{+0.5}_{-0.7} \times 0.08^{+0.67}_{-0.08}\masmas$, corresponding to linear sizes of
$7.3^{+3.4}_{-7.3} \times 2.7^{+4.1}_{-2.7}\pcpc$ and $3.7^{+1.1}_{-1.6} \times 0.2^{+1.5}_{-0.2}\pcpc$, 
at 2.3 and 8.4\,GHz, respectively. 
Applying these values, we estimate the intrinsic brightness temperatures, $T_\mathrm{b}'$, as $\sim 10^{8.0}$\,K at 2.3\,GHz and $\sim 10^{7.8}$\,K at 8.4\,GHz.

\subsection{F01298\texorpdfstring{$-$}{--}0744}
Full-resolution images without UV tapering achieved a beam size of $8.4 \times 2.7\masmas$ ($20 \times 6.5\pcpc$) at 2.3\,GHz, and $2.5 \times 1.1\masmas$ ($6.0 \times 2.6\pcpc$) at 8.4\,GHz (Table\,\ref{tbl:image}). 
In the images at both frequencies, we observed no emission exceeding the 5$\sigma$ threshold of thermal noise.
\cite{2024A&A...687A.193W} reported the detection of 1.5\,GHz continuum and OH megamaser emission through the VLBA, associated with F01298$-$0744.
They also reduced the same data with us and have identified 8.4\,GHz emission with a 4$\sigma$ significance in the full-resolution image, potentially suggesting the presence of a genuine feature in the source.

\begin{figure*}[ht]
    \centering
    \includegraphics[width=0.8\linewidth]{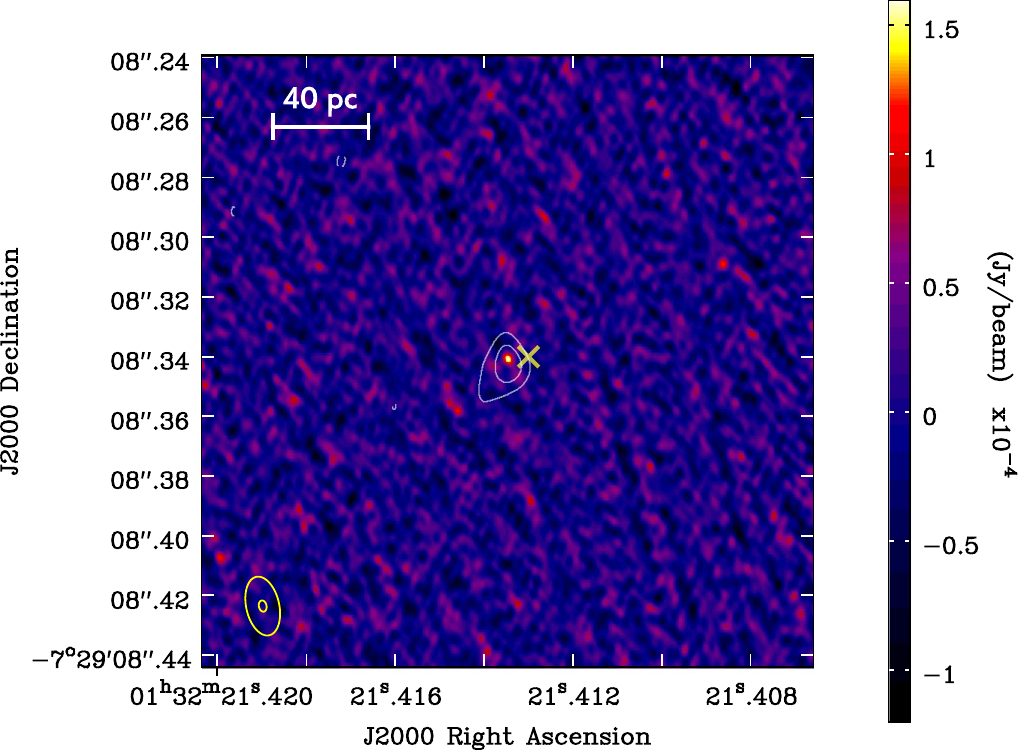}
    \caption{
        A naturally weighted VLBA image of F01298$-$0744 at 8.4\,GHz with 2.3 GHz contours overlaid.
        The UV data are tapered by a Gaussian with a FWHM of 10\,M$\lambda$ for 2.3\,GHz and 50\,M$\lambda$ for 8.4\,GHz.
        Contour levels begin at 1212\uJybm, corresponding to the 3$\sigma$ noise level of the image, and increase by factors of $\sqrt{2}$. 
        The resolutions at 2.3 and 8.4\,GHz are 20.02$\times$11.04 mas$^2$ at a PA of 12\fdg6 and 
        3.79$\times$2.56 mas$^2$ at a PA of 11\fdg8, 
        respectively, indicated by the yellow ellipses in the bottom left of the image.
        A yellow cross indicates the location of the submillimeter continuum source detected by ALMA observations \citep{2019ApJS..241...19I}, which corresponds to the phase center in imaging process.
        The peak flux densities are 2081\uJybm and 163\uJybm at 2.3 and 8.4\,GHz, respectively.
        \label{fig:F01298img}
        }
\end{figure*}

To further assess the significance of the result presented by \cite{2024A&A...687A.193W}, we produced low-resolution images with UV tapering.
Figure\,\ref{fig:F01298img} shows the UV tapered image of F001298$-$0744 without self-calibration.
The flux peak is located at
$\mathrm{R.A.} = 01^{\mathrm h}32^{\mathrm m}21\fs41340 \pm 0\fs00007$ and 
$\mathrm{decl.} = -07\arcdeg29\arcmin08\farcs343 \pm 0\farcs003$ at 2.3\,GHz, and 
$\mathrm{R.A.} = 01^{\mathrm h}32^{\mathrm m}21\fs41348 \pm 0\fs00003$ and 
$\mathrm{decl.} = -07\arcdeg29\arcmin08\farcs341 \pm 0\farcs001$ at 8.4\,GHz.
See Appendix \ref{sec:position} for the details of the position error estimates.
The thermal noise and peak flux density measured in the image are 404 and 2081\uJybm at 2.3\,GHz and 27 and 201\uJybm at 8.4\,GHz, respectively.
The positional offset of the radio sources between the two frequencies amounts to 1.8$\sigma$.
The specific intensity exceeds the 5$\sigma$ threshold at the same position within the error margin in the 2.3 and 8.4\,GHz images.
Thus, this radio source is unlikely to result from noise.
The Gaussian fit for the image performed using the AIPS task IMFIT yields a FWHM of the source as $21\pm 3 \times 20 \pm 3\masmas$ on the 2.3\,GHz image, whereas $3.8\pm 0.6 \times 3.0 \pm  0.5\masmas$ on the 8.4\,GHz image, which are more extended than the beam sizes. 
We estimated the flux density of the source by summing the CLEAN components within these regions, yielding a value of 2740 and 201\,$\mu$Jy at 2.3 and 8.4\,GHz, respectively (Table\,\ref{tbl:flux}).

Based on the source size in the images, the brightness temperature of the source is estimated as $T_\mathrm{b}\sim10^{6.3}$\,K at 2.3\,GHz and $T_\mathrm{b}\sim10^{5.7}$\,K at 8.4\,GHz using Equation\,(\ref{eq:Tb}).
The deconvolved size at 2.3\,GHz for the compact radio source detected was measured as 
$16^{+5}_{-16} \times 7^{+13}_{-7}\masmas$, corresponding to the linear size of $39^{+12}_{-39} \times 17^{+31}_{-17}\pcpc$, by the IMFIT.
On the other hand, the only measurable quantity at 8.4\,GHz was the major axis, which was $1.6^{+1.0}_{-1.6}$\,mas, corresponding to $3.9^{+2.4}_{-3.9}$\,pc, while the minor axis yielded an upper limit of 1.6\,mas, corresponding to 3.9\,pc.
Using these values and the integrated flux densities, the intrinsic brightness temperature can be further estimated as $T_\mathrm{b}'\sim 10^{7.0}$\,K at 2.3\,GHz and $T_\mathrm{b}' > 10^{6.1}$\,K at 8.4\,GHz.

\section{Discussion} \label{sec:discuss}

\subsection{Origin of the radio sources detected by the VLBA}
\subsubsection{F00188\texorpdfstring{$-$}{--}0856}
The compact radio source in F00188$-$0856 exhibits an high intrinsic brightness temperature of $T_\mathrm{b}'\sim10^{7.8}$\,K at 8.4\,GHz. 
This signifies the presence of nonthermal processes, such as an AGN, a radio supernova (RSN), or a supernova remnant (SNR), given that compact starbursts devoid of such phenomena exhibit brightness temperatures of $T_{\mathrm b}' \lesssim 10^{5}$\,K \citep[][]{1992ARA&A..30..575C}.
Empirical correlations, known as the $\Sigma$-$D$ relation, establish a relationship between the surface brightness and the physical size of RSNe and SNRs \citep{1976MNRAS.174..267C,1985ApJ...295L..13H}. 
A 1-pc RSN or SNR typically attains a surface brightness of $\sim10^{-15}$\,W\,m$^{-2}$\,Hz$^{-1}$\,str$^{-1}$ \citep[e.g.,][]{2005A&A...435..437U,2010A&A...509A..34B}, corresponding to $T_{\mathrm b}'\sim10^{6}$\,K. Consequently, a single RSN or SNR cannot account for the observed features.
Although multiple SNRs/RSNe densely clustered within a $\sim1$\,pc region could potentially reproduce the high intrinsic brightness temperature, it remains highly implausible, given the typical separation of $\sim10$\,pc between RSNe/SNRs in the nucleus of Arp\,220 \citep[e.g.,][]{2006ApJ...647..185L}. Thus, achieving $T_\mathrm{b}'\sim10^{7.8}$\,K is only possible through AGN activity.

At X-ray, a point source has been identified at the coordinates of
$\mathrm{R.A.}=00^\mathrm{h}21^\mathrm{m}26\fs54$, 
$\mathrm{DEC.}=-08\arcdeg39\arcmin25\farcs9$
obtained through the
Chandra X-ray Observatory \citep{2005ApJ...633..664T} using the Advanced CCD Imaging Spectrometer (ACIS)\footnote{The reference paper does not provide a position error, which would typically be on the order of 0.1\,arcseconds. 
This estimate is based on the assumption that the ACIS observations with 1-arcsecond spatial resolution detect a signal greater than 5$\sigma$.}.
The position of the compact radio source aligns with that of the X-ray source, whose hardness ratio indicates its soft spectrum, suggesting a Compton-thick circumnuclear environment. 
This argument aligns with indications of AGN obscuration in F00188$-$0856 \citep{2007ApJS..171...72I,2008PASJ...60S.489I,2009ApJS..182..628V,2010MNRAS.405.2505N}.

F00188$-$0856 exhibits a far-infrared-to-radio ratio consistent with the mean value (H21), within the intrinsic scatter observed for entire galaxies \citep{2001ApJ...554..803Y}. 
Although a radio-loud AGN typically amplifies radio emission, F00188$-$0856, despite hosting an AGN, lacks this signature. 
Recent studies have identified AGNs even in systems with moderate far-infrared-to-radio ratios \citep[e.g.,][]{2010ApJ...724..779M,2019PASJ...71...28S}, with F00188$-$0856 serving as a compelling example.

\subsubsection{F01298\texorpdfstring{$-$}{--}0744}
\label{sec:dis-01298}
In F01298$-$0744, \cite{2024A&A...687A.193W} have reported a radio source detected with the VLBA at 1.5\,GHz, exhibiting a integrated flux density of $\sim2$\,mJy and an angular extent of $50\times20\masmas$ in the 1.5\,GHz image, corresponding to a brightness temperature of $\sim10^{6.5}$\,K.
We also identified this source in the tapered image at 2.3 and 8.4\,GHz, constructed by weighting the short baselines. 
The inferred intrinsic brightness temperature exceeded $10^{6.1}$\,K at 8.4\,GHz, consistent with the findings of \cite{2024A&A...687A.193W}.
While these values exceed the upper limit for a starburst without RSNe or SNRs \citep[$\lesssim 10^5$\,K;][]{1992ARA&A..30..575C}, a brightness temperature of  $\sim10^6$\,K remains consistent with emission from a RSN or SNR within the detected sources. 
Although the source suggests an intrinsic brightness temperature of $10^{7.0}$\,K at 2.3\,GHz (Table\,\ref{tbl:flux}), flux measurements at this frequency suffer substantial uncertainty due to the severe RFI (Section\,\ref{sec:reduction}).
This compact radio emission may originate from an AGN, but star formation containing a RSN or SNR still offers a viable alternative explanation \citep[see][]{2024A&A...687A.193W}.
Therefore, the present observations do not provide compelling evidence for AGN activity in F01298$-$0744. 
Assessing whether their flux variability exhibits signatures of an RSN or SNR remains essential \citep[e.g.,][]{2019A&A...623A.173V}.

If the compact radio source originates from starburst activity, the star formation rate within the source (in $M_\odot\,\mathrm{yr}^{-1}$) is expressed as 
\begin{eqnarray} 
\mathrm{SFR} = 258.4 \times 10^{-30} L_\mathrm{8.4}, \label{eq:sfr} 
\end{eqnarray} 
where $L_\mathrm{8.4}$ represents the specific radio luminosity at 8.4\,GHz in units of \ergsHz \citep{2003A&A...409...99P,2010MNRAS.405..887C}. 
As a result, the star formation rate is determined to be 25\,$M_\odot\,\mathrm{yr}^{-1}$, based on the specific luminosity at 8.4\,GHz of $9.5\times 10^{28}$\ergsHz.
The core-collapse supernova rate is approximately 0.01 times the star formation rate \citep[e.g.,][]{2012A&A...537A.132B,2012A&A...545A..96M}, assuming a Salpeter initial mass function \citep{1955ApJ...121..161S}.
Consequently, the core-collapse supernova rate is estimated to be $\sim0.3\,\mathrm{yr}^{-1}$ for the nucleus of F01298$-$0744.




\subsection{Mid-Infrared and Radio Diagnostics of Obscured AGNs in ULIRGs}
The three ULIRGs examined in this study exhibit mid-infrared characteristics indicative of an obscured AGN \citep{2006ApJ...637..114I,2007ApJS..171...72I}.
One of the objectives of this study is to confirm the presence of such an AGN through high-resolution radio observations.
We detected a compact radio source with a high brightness temperature, unambiguously attributable to an AGN, in F00188$-$0856, for which
the infrared diagnosis is shown to be correct.
In contrast, although the compact radio source identified in F01298$-$0744 may originate from an AGN, conclusive evidence has not been found to support the existence of AGNs in either of the remaining two objects, F00091$-$0756 and F01298$-$0744.
These results indicate that VLBI-scale radio features of ULIRGs do not necessarily corroborate the presence of buried AGNs.
However, this finding does not invalidate mid-infrared diagnostics.
Although the bolometric contribution of AGNs in our targets is about 30--60\% \citep{2010MNRAS.405.2505N}, the compact radio source in F00188$-$0856 containing a radio AGN only accounts for 10\% of the integrated flux density.
The radio emission from AGNs in ULIRGs may be less significant than their bolometric output, or the bolometric contribution of AGNs may be overestimated \citep{2024MNRAS.529.4468L}.
Furthermore, if the spatial resolution or sensitivity of the observations is insufficient to resolve radio emission, AGNs may display brightness temperatures below $10^6$\,K, comparable to those of starburst regions \citep[e.g.,][]{2022ApJ...940...52S}.
A systematic survey of nearby ULIRGs, including sources like F00091$-$0756 and F01298$-$0744, using high-sensitivity, high-resolution instruments such as the VLBI capabilities of the Square Kilometer Array and/or the ngVLA is essential to advance the study of radio AGNs in ULIRGs.

\subsection{Spatial Distribution of Radio Continuum Emission}
\label{sec:extended}

All sources analyzed in this study exhibit integrated flux densities of 3--4\,mJy in previous VLA observations at 9.0\,GHz (Table\,\ref{tbl:sample}), likely corresponding to comparable values at 8.4\,GHz. 
The previous VLA data, obtained in the C configuration with an angular resolution of $\sim2\arcsec$, show that all the targets are unresolved (H21).
VLBA observations in this study detected compact radio emission in two of the three targets, with 8.4\,GHz flux densities not exceeding 300\,$\mu$Jy, contributing merely $\sim10$\% to the integrated flux density measured by the VLA. 
The shortest baseline of the VLBA, which extends several hundred kilometers, constrains the maximum recoverable angular scale to $\sim$10\,mas at 8.4\,GHz.
Thus, the emission resolved out by the VLBA but detected by the VLA which is 3--4\,mJy originates from intermediate spatial scales of $\sim10$\,mas to $\sim2\arcsec$, corresponding to the physical extents of $\sim20$\,pc to $\sim4$\,kpc.

\subsection{Origin of the Spectral Steepening}
Now, we evaluate whether compact radio sources identified by the VLBA or diffuse extended emission undetected by the VLBA is responsible for the spectral steepening at $\sim10$\,GHz observed with the VLA (H21). 
Then, we investigate astrophysical processes driving these phenomena.

\subsubsection{Case for the Compact Radio Sources}
\label{sec:case-compact}
Assuming that extended emission exhibits a steep power-law spectrum without any high-frequency steepening and that a compact source as identified in this study, which contributes 10\% to the integrated flux density at $\lesssim10$\,GHz, is not detectable at $\gtrsim10$\,GHz due to its faintness, the spectral index between 9 and 14\,GHz, derived from arcsecond-resolution observations, would decrease by 0.2 relative to that measured at lower frequencies.
Thus, a pronounced spectral steepening of the compact radio source could explain the observed spectral curvature in the VLA data. 
This exponential steepening, inferred from spectral modeling of arcsecond-scale radio observations \citep[e.g.,][]{2018A&A...611A..55K}, likely results from the instantaneous injection of accelerated particles \citep{1962SvA.....6..317K,1973A&A....26..423J}.
In a compact AGN-driven radio source such as F00188$-$0856, continuous particle injection is expected, making significant spectral steepening due to instantaneous injection improbable.
We therefore assess the plausibility that a compact radio source powered by starburst activity, such as F01298$-$0744, accounts for the observed spectral steepening, leveraging the core-collapse supernova rate estimated in Section\,\ref{sec:dis-01298}.

The spectral age of the source, $t_\mathrm{s}$ (in yr), is expressed as
\begin{eqnarray} 
t_\mathrm{s} = 
1.61\times 10^9
\frac{B^{\frac{1}{2}}}{ B^2 + B_\mathrm{iC}^2} 
\nu_\mathrm{b}^{-\frac{1}{2}}, 
\label{eq:sync-age} 
\end{eqnarray}  
where $\nu_\mathrm{b}$ denotes the rest-frame spectral break frequency (in GHz), $B$ is the magnetic field strength (in $\mu$G), and $B_\mathrm{iC}$ represents the effective magnetic field corresponding to the radiation energy density, accounting for inverse Compton losses of relativistic particles (in $\mu$G).
$B_\mathrm{iC}$ includes the contributions from the far-infrared photon field, estimated from the infrared bolometric luminosity and spatial extent of submillimeter continuum sources \citep{2019ApJS..241...19I}, as well as the cosmic microwave background, $3.25(1+z)^2\,\mu$G \citep[e.g.,][]{1999A&A...345..769M,2021MNRAS.503.5746N}.
Note that Equation (\ref{eq:sync-age}) remains identical regardless of whether the accelerated particles are injected instantaneously or continuously (see \citealt{1991ApJ...383..554C,2003PASA...20...19M} and references therein).

Under the assumption of minimum energy condition, where the energy densities of relativistic electrons and magnetic fields are approximately equal, the equipartition magnetic field strength, $B_\mathrm{eq}$ (in $\mu\mathrm{G}$), is calculated as
\begin{eqnarray} 
B_\mathrm{eq} = 
2.9\times10^{-8} 
(1+k)^{\frac{2}{7}}
L_\mathrm{bol}^{\frac{2}{7}} 
R^{-\frac{6}{7}} 
, 
\label{eq:B_eq} 
\end{eqnarray} 
where $k$, $L_\mathrm{bol}$, and $R$ correspond to the the energy ratio between heavy particles and electrons, the bolometric radio luminosity (in \ergs), and source size (in pc) \citep{1970ranp.book.....P}.
We adopt $k = 40$ \citep{1991ASPC...18...37W}, but note that $B_\mathrm{eq}$ has minimal sensitivity to variations in $k$.
Here, the bolometric luminosity, $L_\mathrm{bol}$, is determined by integrating over the frequency range of 10\,MHz to 100\,GHz, assuming a spectral index of $\alpha=-0.7$.
For the compact radio source characterized by flux densities of 200\,$\mu$Jy and deconvolved spatial extents of a few pc at $z = 0.136$, corresponding to $L_\mathrm{bol} = 2 \times 10^{39}$\ergs, as observed in F01298$-$0756 (see Table\,\ref{tbl:flux}), we estimate $B_\mathrm{eq}$ to lie within the range of $\sim1$--10\,mG.

Setting $B = B_\mathrm{eq}$, the resulting $t_\mathrm{s}$ for compact radio sources exhibiting spectral steepening near 10\,GHz, as observed in F01298$-$0744, falls within $\sim 10^2$--$10^3$\,yr. 
In this regime, synchrotron losses predominate over inverse Compton losses.
If the compact radio source originates from starburst activity, the core-collapse supernova rate, estimated as $\sim0.3\,\mathrm{yr}^{-1}$ for F01298$-$0744 (Section\,\ref{sec:dis-01298}), renders the assumption of instantaneous injection --- implying negligible particle acceleration over $\sim10^2$--$10^3$\,yr --- highly untenable.
Thus, compact radio sources, as identified in this study, whether associated with AGNs or starbursts, are unlikely to explain the spectral steepening observed in the past VLA observations.


\subsubsection{Case for the Extended Emission}
\label{sec:case-extended}
In contrast to the previous argument, extended emission, undetected by the VLBA but detected by the VLA, induces spectral steepening around 10\,GHz. 
The subsequent analysis evaluates whether its spectral age aligns with astrophysical expectations.

\begin{figure*}[ht!]
    \epsscale{0.55}
    \plotone{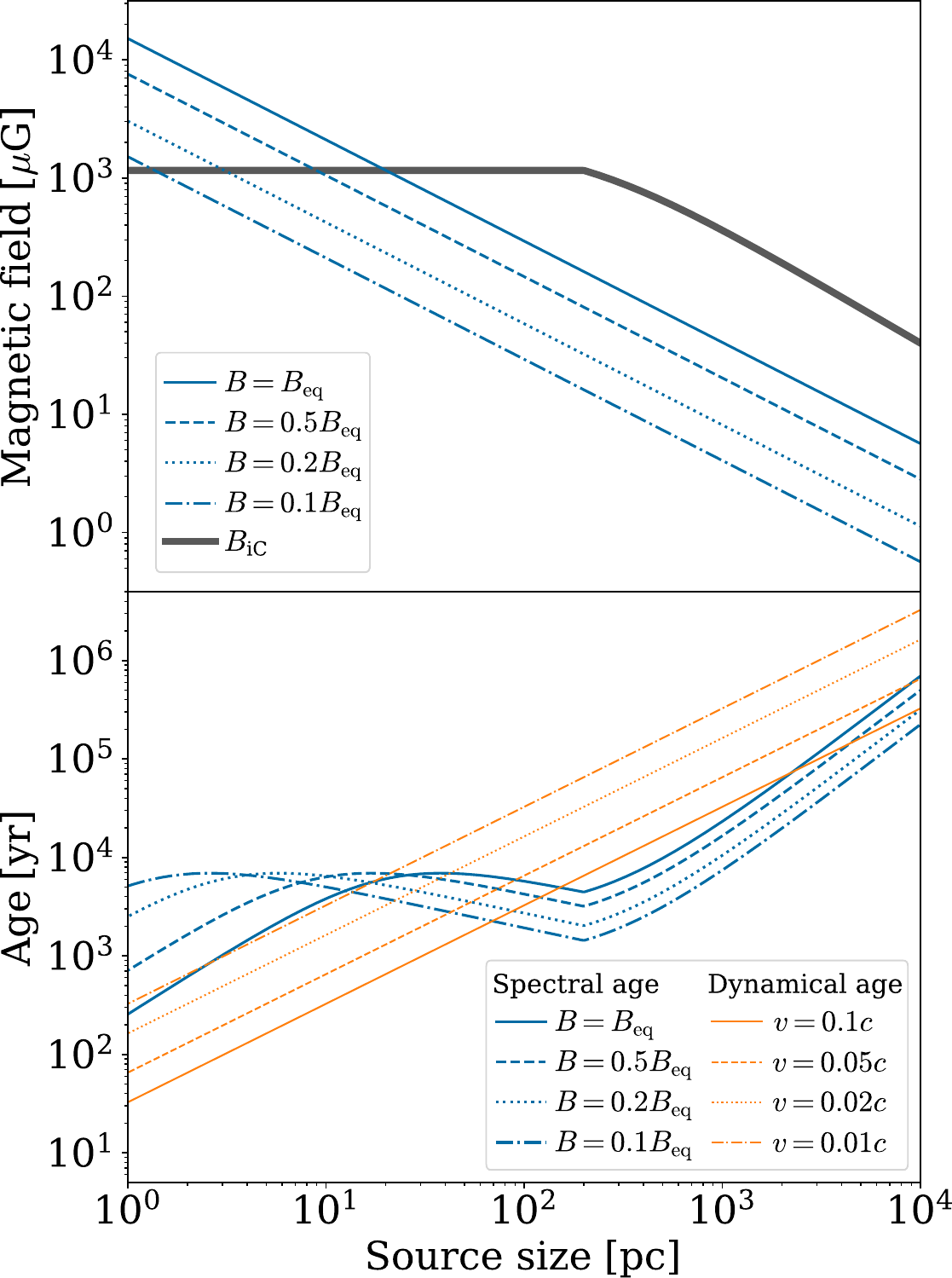}
    \caption{
    The magnetic field strength (top) and age (bottom) of nonthermal plasma, resolved out by the VLBA but unresolved by the VLA, as functions of the assumed source size, $R$.
    We consider sources at a redshift of 0.13, exhibiting a 8.4\,GHz flux density of 3\,mJy and a spectral index of $-0.7$.
    Note that this figure does not illustrate the evolution of the source but instead presents the inferred physical conditions associated with each assumed source size, which remains observationally unconstrained.
    Blue solid, dashed, dotted, and dash-dotted lines in the top panel represent cases where the magnetic field strength, $B$, is 1, 0.5, 0.2, and 0.1 times the equipartition value, $B_\mathrm{eq}$ (Equation\,(\ref{eq:B_eq})), respectively.
    The corresponding spectral age, $t_\mathrm{s}$, inferred assuming a spectral break at $\nu_\mathrm{b}=10$\,GHz as observed in H21, is shown by the blue lines in the bottom panel, incorporating synchrotron and inverse Compton losses (Equation\,(\ref{eq:sync-age})).
    Orange solid, dashed, dotted, and dash-dotted lines in the bottom panel depict the dynamical age, $t_\mathrm{dyn}$, assuming expansion velocities, $v$, of $0.1c$, $0.05c$, $0.02c$, and $0.01c$, respectively.
    A bold gray line in the top panel indicates the effective magnetic field, $B_\mathrm{iC}$, equivalent to the photon energy density for inverse Compton losses.
    The radiation field is modeled as a sphere with a 150\,pc radius \citep[see][]{2019ApJS..241...19I}, encompassing uniformly distributed radiant energy assuming a bolometric infrared luminosity of $10^{12.3}L_\odot$.
    The external energy density, scaling as the inverse square of distance, is averaged over 150\,pc to the source's outer boundary.
    There are three distinct regimes in the bottom panel: 
    the synchrotron-dominated regime with constant $B_\mathrm{iC}$ ($R\lesssim 10^1$\,pc; rising to the right), 
    the inverse Compton-dominated regime with constant $B_\mathrm{iC}$ ($10^1\,\mathrm{pc} \lesssim R\lesssim 10^2$\,pc; falling to the right), 
    and the inverse Compton-dominated regime where $B_\mathrm{iC}$ decreases as $R$ increases ($10^2\,\mathrm{pc} \lesssim R$; rising to the right).
    }
    \label{fig:age-R}
\end{figure*}

As in Section\,\ref{sec:case-compact}, we estimate the spectral age of the extended emission, $t_\mathrm{s}$, using Equation\,(\ref{eq:sync-age}). 
The magnetic field strength, $B$, necessary for this calculation follows from the equipartition condition given by Equation\,(\ref{eq:B_eq}). 
Given the substantial uncertainty in the spatial extent of the extended emission --- a key parameter for determining the equipartition magnetic field strength, $B_\mathrm{eq}$ --- which spans a wide range from 20\,pc to 4\,kpc (Section\,\ref{sec:extended}), 
we assess each assumed spatial scale individually.
The corresponding physical constraints appear in Figure\,\ref{fig:age-R}.
Note that this figure does not illustrate the evolutionary trajectory of the radio sources. 
The spectral age exhibits a non-monotonic dependence on the assumed source size, reflecting the varying predominance of inverse Compton and synchrotron cooling.
As a result, for source sizes between 20\,pc and 4\,kpc, 
we constrain the spectral age and magnetic field strength to $\sim10^4$--$10^5$\,yr and $\sim10^1$--$10^3\,\mu$G, respectively.

One compelling scenario that accounts for the estimated parameters mentioned above involves merger-driven nonthermal processes.
\cite{2013ApJ...777...58M} have reported that as galaxy mergers progress, radio bridges and tidal tails with steep spectra at high frequencies emerge, undergoing rapid spectral aging in a magnetized medium shaped by dynamical interactions \citep[see also][]{1993AJ....105.1730C,2002AJ....123.1881C}.
The radio emission associated with this phenomenon typically extends over several kpc, exhibits magnetic field strengths on the order of 10\,$\mu$G, and suggests electron cooling timescales of $\sim 10^4$\,yr \citep{2013ApJ...777...58M}. 
The extended emission from all of our targets satisfies these physical conditions. 
This scenario is particularly applicable to F00091$-$0738 and F01298$-$0744, which reside in the pre-merger or active merger stages \citep{2002ApJS..143..315V}.
The far-infrared-to-radio ratios of both sources are typical or slightly lower than average (H21), consistent with a merger-driven origin of the extended emission \citep{2013ApJ...777...58M}.

Although F00188$-$0856 exhibits neither multiple nuclei, prominent tidal features, nor morphological distortions in the optical regime
--- consistent with its classification as a late-stage merger \citep{2002ApJS..143..315V} --- subtle signatures of prior dynamical interactions may persist. 
The extended radio emission could therefore arise from merger-driven particle acceleration, analogous to the processes inferred in the other two systems. 
Alternatively, the observed large-scale emission may reflect radio lobes powered by AGN jets launched from the compact, AGN-dominated VLBI core, which displays high-frequency spectral steepening. 
A similar interpretation can apply to F01298$-$0744, whose VLBI-scale emission may originates from an AGN as well as starburst activity. 
Note that, in these cases, the relationship between merger activity and radio emission remains indeterminate.

To assess the viability of an AGN origin of the extended emission, we examined whether the dynamical age of the lobe, $t_\mathrm{dyn}$, inferred from the typical lobe advance speed, aligns with the spectral age, $t_\mathrm{s}$. 
Given that the expansion velocity of the lobe, $v$, does not exceed $\sim0.1c$ \citep{2008ApJ...687..141K}, the dynamical and spectral ages of the proposed lobe are comparable if the source size is $\lesssim 100$\,pc or $\gtrsim 1$\,kpc, corresponding to estimated ages of $\sim10^4$\,yr and $\sim10^5$\,yr, respectively (Figure \ref{fig:age-R}).
Thus, AGN activity can be the origin of a spectral steepening of the extended emission.
The substantial nonthermal fraction (H21) and the presence of a compact radio source with a high intrinsic brightness temperature further reinforce the AGN origin of its extended radio emission.

\begin{figure*}[ht!]
    \epsscale{0.6}
    \plotone{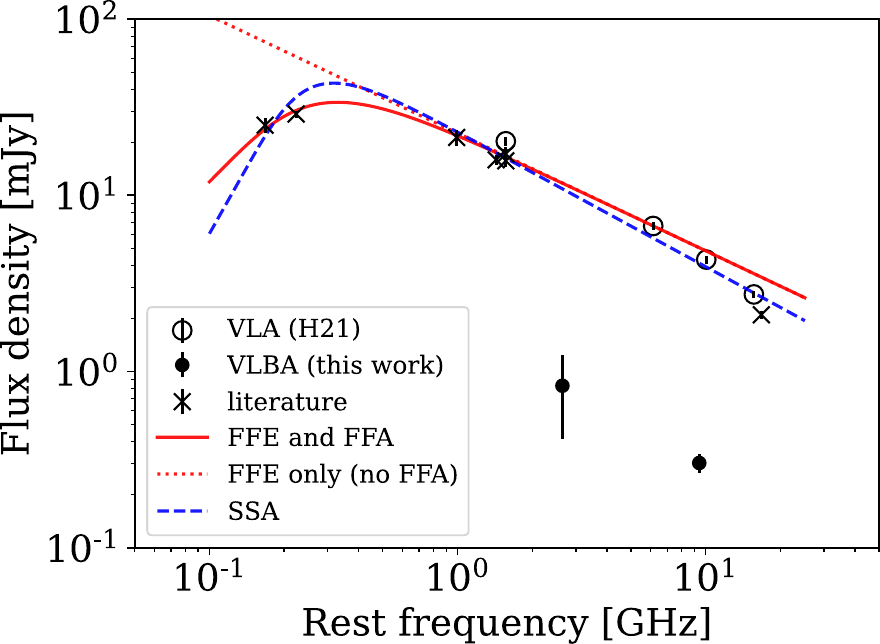}
    \caption{
    Radio spectrum of F00188$-$0856.
    Open circles represent flux densities by VLA measurements at 1.4, 5.5, 9.0, and 14.0\,GHz (H21).
    Crosses denote the integrated flux densities obtained from 
    the GLEAM-X survey at 200\,MHz \citep{2022PASA...39...35H,2024PASA...41...54R}, 
    RACS at 887.5\,MHz \citep{2021PASA...38...58H}, 
    GMRT observations at 150\,MHz \citep[TGSS;][]{2017AA...598A..78I} and 1.28\,GHz \citep{2021MNRAS.503.5746N}, and VLA measurements at 1.4\,GHz (the FIRST survey \citep{1995ApJ...450..559B} and NVSS \citep{1998AJ....115.1693C}) and 15.0\,GHz \citep{2003AA...409..115N}.
    See Table\,\ref{tbl:flux-00188} for details of these measurements.
    Red solid and dotted lines represent a model that incorporates a nonthermal component subject to FFA by an ambient thermal plasma with FFE (Equation\,(\ref{eq:model-FFA})), and an identical model without absorption effects, respectively. 
    A blue dashed line shows a model with a nonthermal component that is affected by SSA (Equation\,(\ref{eq:model-SSA})).
    Flux densities at 2.3 and 8.4\,GHz, derived from the VLBA observations in this study, are marked by filled circles but are excluded from the model fit.
    \label{fig:spectrum}
    }
\end{figure*}

Figure\,\ref{fig:spectrum} shows the radio spectrum of F00188$-$0856, exhibiting a convex profile that peaks near 300\,MHz \citep{2021MNRAS.503.5746N}. 
As summarized in Table\,\ref{tbl:flux-00188}, low-frequency observations exhibit low spatial resolution and preferentially detect diffuse extended emission, yielding systematically higher flux densities than high-frequency measurements with high angular resolution. 
However, this potential bias does not influence the presence of low-frequency absorption features in F00188$-$0856.
The spectral characteristic of F00188$-$0856 identifies it as a young radio source \citep{1998PASP..110..493O,2009AN....330..120F}. 
A well-established correlation links the peak frequency with the physical extent of young radio sources \citep{1990A&A...231..333F,1997AJ....113..148O}. 
F00188$-$0856 follows this relation if its linear size is $\sim 1$\,kpc, in agreement with our estimate in Section\,\ref{sec:case-extended} (Figure\,\ref{fig:age-R}). 
Thus, as \cite{2021MNRAS.503.5746N} noted, F00188$-$0856 hosts a young radio source, which plausibly accounts for the extended emission and the pronounced spectral steepening at high frequencies.

\begin{deluxetable}{cccccccccccc}
\tabletypesize{\scriptsize}
\tablewidth{0pt} 
\tablecaption{Flux Measurement of F00188$-$0856 \label{tbl:flux-00188}}
\tablehead{													
\colhead{Frequency}	&	\colhead{Flux density}			&	\colhead{Beam size}			&	\colhead{PA}	&	\colhead{Reference}	\\
\colhead{(MHz)}	&	\colhead{(mJy)}			&	\colhead{(arcsec$^2$)}			&	\colhead{}	&	\colhead{}	\\
}													
\startdata 													
\phn\phn150	&	25.0	$\pm$	2.5	&	25 	$\times$	25 	&	0	&	(1)	\\
\phn\phn200	&	28.9	$\pm$	1.3	&	76 	$\times$	57 	&	$-30$	&	(2)	\\
\phn\phn888	&	21.3	$\pm$	2.2	&	25 	$\times$	25 	&	0	&	(3)	\\
\phn1280	&	15.9	$\pm$	1.1	&	4.8 	$\times$	4.8 	&	0	&	(4)	\\
\phn1400	&	17.2	$\pm$	1.6	&	5 	$\times$	5 	&	0	&	(5)	\\
\phn1400	&	15.7	$\pm$	1.2	&	45 	$\times$	45 	&	0	&	(6)	\\
\phn1400	&	20.3	$\pm$	1.2	&	7.8 	$\times$	3.9 	&	141	&	(7)	\\
\phn5500	&	6.70	$\pm$	0.34	&	5.0 	$\times$	3.5 	&	173	&	(7)	\\
\phn9000	&	4.32	$\pm$	0.22	&	3.5 	$\times$	2.0 	&	149	&	(7)	\\
14000	&	2.74	$\pm$	0.14	&	9.6 	$\times$	4.4 	&	135	&	(7)	\\
15000	&	2.1	$\pm$	0.1	&		0.15   $\times$   0.15\tablenotemark{a}		&	...	&	(8)	\\
\enddata													  				
\tablecomments{
    References:
    (1) \cite{2017AA...598A..78I}, 
    (2) \cite{2022PASA...39...35H,2024PASA...41...54R}, 
    (3) \cite{2021PASA...38...58H}, 
    (4) \cite{2021MNRAS.503.5746N}, 
    (5) \cite{1995ApJ...450..559B}, 
    (6) \cite{1998AJ....115.1693C}, 
    (7) H21, 
    (8) \cite{2003AA...409..115N}
}
\tablenotetext{a}{\cite{2003AA...409..115N} presents a survey observations of ULIRGs, reporting only overall spatial resolution of their sample ($\sim$150\,mas) without specifying that of F00188$-$0856.}
\end{deluxetable}

\subsection{Impact of AGN activity in F00188\texorpdfstring{$-$}{--}0856}
The following analysis examines whether the jet activity that powers the young radio source in F00188$-$0856 influences the galaxy evolution.
Under the equipartition condition, the time-averaged kinetic power of AGN jets, $P_\mathrm{jet}$ (in \ergs), fueling the extended features in a radio galaxy is derived as
\begin{eqnarray}
P_\mathrm{jet}
= 3.4\times10^{14}
f^{\frac{3}{2}}
L_{151}^{\frac{6}{7}},
\label{eq:Pkin_W99}
\end{eqnarray}
where $f$ and $L_{151}$ represent a parameter accounting for systematic error in the model assumptions and the specific radio luminosity at 151\,MHz (in \ergsHz), respectively \citep{1999MNRAS.309.1017W}.
This relationship applies regardless of the radio power of the jet activities \citep{2013ApJ...767...12G}.
Due to significant absorption at low frequencies, the 150\,MHz flux density reported in \cite{2021MNRAS.503.5746N}, derived from the TIFR GMRT Sky Survey \citep[TGSS;][]{2017AA...598A..78I}, is not suitable for computing $L_{151}$. 
Instead, we estimate the unabsorbed intrinsic flux density at this frequency.
A model incorporating a single nonthermal component with free-free absorption (FFA), 
\begin{eqnarray}
S_\nu \propto \bigl(1-\exp({-\tau_\nu})\bigl) \Bigl[ 1+\frac{1}{H}\Bigl(\frac{\nu}{\mathrm{GHz}}\Bigl)^{0.1+\alpha} \Bigl]\nu^2,
\label{eq:model-FFA}
\end{eqnarray}
or synchrotron self-absorption (SSA),  
\begin{eqnarray}
S_\nu \propto \nu^{5/2} \Bigl[ 1 - \exp (-\tau_\mathrm{s}\nu^{\alpha-5/2}) \Bigl],
\label{eq:model-SSA}
\end{eqnarray}
is fit to the data points shown in Figure\,\ref{fig:spectrum}.
Here, $\tau_\nu \propto \nu^{-2.1}$ and $\tau_\mathrm{s}$ denote the optical depths attributable to FFA and SSA, respectively, while $H$ represents the ratio of thermal to nonthermal emission components \citep{1991ApJ...378...65C}.
As a result, FFA better accounts for the low-frequency absorption.
Applying the same parameters while excluding the absorption effect due to FFA (i.e., $\tau_\nu\rightarrow 0$), the unabsorbed flux density at 151\,MHz is estimated to be $\sim80$\,mJy. 
Using this value, we derive $L_{151} = 3 \times 10^{31}$\ergsHz. 
Note that this estimate remains tentative as a multi-component model incorporating FFA and/or SSA may be relevant.
For a value of $f=20$ applicable to this luminosity range \citep{2013ApJ...767...12G}, 
we obtain $P_\mathrm{jet} \sim 3\times10^{43}$\,\ergs from Equation\,(\ref{eq:Pkin_W99}).
This value is comparable to that of F01004$-$2237, which hosts large-scale radio lobes that span a projected linear extent of 100\,kpc
and manifests a jet power of $P_\mathrm{jet} \sim 3\times10^{43}$\,\ergs \citep{2024ApJ...970....5H}.

F00188$-$0856 shows a broad blueshifted component with $\mathrm{FWHM} \sim 900$\kms and a velocity offset of $\sim -250$\kms in the [\ion{O}{3}] emission line \citep{2013MNRAS.432..138R}, accompanied by a molecular outflow characterized by a density-weighted velocity of $\sim -300$\kms in the OH doublet at 119\,$\mu$m \citep{2013ApJ...775..127S}, indicative of a thermal wind in this system. 
These values are lower than those reported for F01004$-$2237 in the same references. 
Given that the jet kinetic power in F01004$-$2237 surpasses that of the thermal outflow and can expel the surrounding medium \citep{2024ApJ...970....5H}, AGN jet activity in F00188$-$0856 is also anticipated to exert a comparable influence. 
Consequently, F00188$-$0856 could thus represent one of the ULIRGs wherein jet activity, driven by merger events, impacts galaxy evolution \citep[e.g.,][]{2022A&A...665L..11P,2023MNRAS.520.5712S,2024MNRAS.530..446H}.


\section{Summary} \label{sec:sum}
We conducted VLBA imaging at 2.3 and 8.4\,GHz of three ULIRGs exhibiting high-frequency spectral steepening in the radio regime and obscured AGN signatures in the mid-infrared. 
While no VLBI-scale radio emission was detected from F00091$-$0738, we identified pc-scale compact radio sources in F00188$-$0856 and F01298$-$0744, contributing only $\sim10$\% of the integrated flux density measured by the VLA. 
The intrinsic brightness temperatures at 8.4\,GHz are $T'_\mathrm{b} \sim 10^{7.8}$\,K for F00188$-$0856 and $T'_\mathrm{b} > 10^{6.1}$\,K for F01298$-$0744.
The high brightness temperature of F00188$-$0856 indicates nonthermal emission inconsistent with typical RSNe or SNRs, suggesting an AGN contribution.
The compact source in F01298$-$0744 is consistent with a starburst origin; however, an AGN scenario remains plausible, rendering its nature and origin enigmatic.
These findings demonstrate that mid-infrared diagnostics of obscured AGNs align with radio observations in at least one case.

An investigation into the origin of high-frequency steepening suggests that if compact radio sources were responsible, their spectra would require an exponential flux attenuation with frequency. 
However, given that compact radio sources encompass either AGNs or RSNe/SNRs, such a scenario is implausible. 
Thus, in all cases, the observed steepening originates from extended kpc-scale emission.
The inferred physical conditions that caused the observed spectral steepening are consistent with merger-induced particle acceleration or radio lobes powered by AGN activity.
In the case of F00188$-$0856 in particular, the peaked radio spectrum identifies a potential young radio source, which suggest that high-frequency steepening can originate from AGN-driven lobe emission.

\begin{acknowledgments}
This research has used the VizieR catalogue access tool, CDS, Strasbourg, France. 
Additionally, we used the NASA/IPAC Extragalactic Database (NED), operated by the Jet Propulsion Laboratory, California Institute of Technology, under contract with the National Aeronautics and Space Administration.
Cosmological calculations are conducted using a calculator provided by \cite{2006PASP..118.1711W}.
The NRAO operating the VLBA is a facility of the National Science Foundation operated under a cooperative agreement by Associated Universities, Inc.
This work used the Swinburne University of Technology software correlator \citep{2011PASP..123..275D}, developed as part of the Australian Major National Research Facilities Program and operated under license.
\end{acknowledgments}

\vspace{5mm}
\facilities{
    VLBA (NRAO)
    }
\software{
    AIPS \citep{2003ASSL..285..109G},
    astropy \citep{2013A&A...558A..33A,2018AJ....156..123A},
    CASA \citep{2007ASPC..376..127M}, 
    Difmap \citep{1997ASPC..125...77S}. 
    }

\appendix
\section{Positional accuracy of the compact radio sources}\label{sec:position}
The position errors for the compact radio sources in F00188$-$0856 and F01298$-$0744 were calculated as the root-sum-square of individual astrometric error components \citep[e.g.,][]{2006A&A...452.1099P,2011Natur.477..185H,2024ApJ...970....5H}.
The error budget for the phase-referencing observations at 2.3 and 8.4\,GHz presented in this paper is summarized in Table\,\ref{tbl:error}.
The positional accuracy of the phase calibrator was obtained from the ICRF3 \citep{2020A&A...644A.159C}.
The accuracy of the flux peak positions for both the target and calibrator was determined by dividing the beam size by the image signal-to-noise ratio.
We also estimated residual errors caused by propagation delays due to the non-dispersive tropospheric medium \citep[e.g.,][]{1999ApJ...524..816R} and the dispersive ionospheric medium \citep[e.g.,][]{1990AJ.....99.1663L}. 
These estimates assumed a zenith angle of 50\arcdeg and a zenith angle difference of 1\arcdeg between the target and calibrator. 
For the non-dispersive troposphere, we adopted a residual of 3\,cm, based on typical zenith delays modeled in the VLBA correlator \citep{1999ApJ...522..157R}.
The total electron content of the dispersive ionosphere was $\sim 2\times10^{17}$\,m$^{-2}$ over the VLBA stations. 
This value was obtained from the US Total Electron Content Product Archive\footnote{https://www.ngdc.noaa.gov/stp/iono/ustec/index.html}, maintained by the National Oceanic and Atmospheric Administration. 
The global ionospheric model derived from GPS satellites has an accuracy of about 10--25\% \citep{1998RaSc...33..565M}. Conservatively, we assumed an uncertainty of 25\%, corresponding to $\sim 5\times10^{16}$\,m$^{-2}$.
Contributions from errors in Earth orientation parameters and antenna positions were estimated following the simulation results presented in \cite{2006A&A...452.1099P}.

\begin{deluxetable*}{lccccccccccccc}
\tabletypesize{\scriptsize}
\tablewidth{0pt} 
\tablecaption{Error budget for the VLBA observations \label{tbl:error}}
\tablehead{																				
\colhead{Target}	&	\multicolumn{5}{c}{F00188$-$0856}								&	&	\multicolumn{5}{c}{F01298$-$0744}								\\\cline{2-6}\cline{8-12}
\colhead{Band}	&	\multicolumn{2}{c}{2.3\,GHz}			&	&	\multicolumn{2}{c}{8.4\,GHz}			&	&	\multicolumn{2}{c}{2.3\,GHz}			&	&	\multicolumn{2}{c}{8.4\,GHz}			\\\cline{2-3}\cline{5-6}\cline{8-9}\cline{11-12}
\colhead{Error component}	&	\colhead{R.A.}	&	\colhead{decl.}	&	&	\colhead{R.A.}	&	\colhead{decl.}	&	&	\colhead{R.A.}	&	\colhead{decl.}	&	&	\colhead{R.A.}	&	\colhead{decl.}	\\
\colhead{}	&	\colhead{($\mu$as)}	&	\colhead{($\mu$as)}	&	&	\colhead{($\mu$as)}	&	\colhead{($\mu$as)}	&	&	\colhead{($\mu$as)}	&	\colhead{($\mu$as)}	&	&	\colhead{($\mu$as)}	&	\colhead{($\mu$as)}	\\
}					
\startdata 				
Source coordinates (calibrator)	&	160	&	340	&	&	160	&	340	&	&	230	&	480	&	&	230	&	480	\\
Flux peak accuracy (calibrator)	&	38	&	116	&	&	4	&	10	&	&	13	&	40	&	&	4	&	10	\\
Flux peak accuracy (target)	&	416	&	1183	&	&	247	&	590	&	&	526	&	1494	&	&	95	&	226	\\
Tropospheric residuals	&	50	&	122	&	&	34	&	97	&	&	50	&	122	&	&	34	&	97	\\
Ionospheric residuals	&	202	&	493	&	&	10	&	29	&	&	202	&	493	&	&	10	&	29	\\
Antenna position	&	2	&	6	&	&	2	&	6	&	&	2	&	6	&	&	2	&	6	\\
Earth orientation	&	3	&	4	&	&	3	&	4	&	&	3	&	4	&	&	3	&	4	\\
Total	&	871	&	2264	&	&	460	&	1076	&	&	1026	&	2639	&	&	378	&	852	\\
\enddata				
\end{deluxetable*}

\bibliography{ULIRG_2024}{}
\bibliographystyle{aasjournal}



\end{document}